\documentclass[10pt]{article}

\usepackage[letterpaper,margin=1in]{geometry}
\usepackage{setspace}
\usepackage[T1]{fontenc}
\usepackage[utf8]{inputenc} 
\usepackage{microtype}
\usepackage{ragged2e}
\usepackage{graphicx} 
\usepackage{tabularx}
\usepackage[hidelinks]{hyperref}
\usepackage{booktabs}
\usepackage{comment}
\usepackage{float}
\usepackage{multirow}
\usepackage{array}
\usepackage{caption}
\usepackage{titlesec}
\titleformat{\section}{\bfseries}{\thesection.}{0.6em}{}
\titlespacing*{\section}{0pt}{8pt}{4pt}
\usepackage[superscript]{cite} 
\newcommand{\authblock}[1]{\textbf{#1}}
\newcommand{\affilblock}[1]{\normalsize #1}

\begin{document}

\begin{center}
{\Large \textbf{Ablation Study of a Fairness Auditing Agentic System for Bias Mitigation in Early-Onset Colorectal Cancer Detection}}\\[6pt]

{\authblock{Amalia Ionescu, PhD\textsuperscript{1,2}; Jose Guadalupe Hernandez, PhD \textsuperscript{1,2}; Jui-Hsuan Chang, MS\textsuperscript{1}; Emily F. Wong, PhD\textsuperscript{1}; Paul Wang, PhD\textsuperscript{1}; Jason H. Moore, PhD\textsuperscript{1}; Tiffani J. Bright, PhD\textsuperscript{1}}}\\[4pt]

{\affilblock{\textsuperscript{1}Cedars-Sinai Medical Center, Los Angeles, CA, USA} \\
\affilblock{\textsuperscript{2}Authors contributed equally in this work}}\\[8pt]
\end{center}

\section*{Abstract}
{\itshape\justifying
Artificial intelligence (AI) is increasingly used in clinical settings, yet limited oversight and domain expertise can allow algorithmic bias and safety risks to persist. This study evaluates whether an agentic AI system can support auditing biomedical machine learning models for fairness in early-onset colorectal cancer (EO-CRC), a condition with documented demographic disparities. We implemented a two-agent architecture consisting of a Domain Expert Agent that synthesizes literature on EO-CRC disparities and a Fairness Consultant Agent that recommends sensitive attributes and fairness metrics for model evaluation. An ablation study compared three Ollama large language models (8B, 20B, and 120B parameters) across three configurations: pretrained LLM-only, Agent without Retrieval-Augmented Generation (RAG), and Agent with RAG. Across models, the Agent with RAG achieved the highest semantic similarity to expert-derived reference statements, particularly for disparity identification, suggesting agentic systems with retrieval may help scale fairness auditing in clinical AI. 
\par}

\section*{Introduction}

Artificial intelligence (AI)–enabled clinical decision support systems are increasingly embedded in routine care, with approximately 70\% of U.S. hospitals integrating predictive machine learning (ML) models into electronic health records (EHRs) to support diagnosis and risk stratification \cite{chang2025}. 
Yet many systems operate with limited transparency and minimal post-deployment monitoring, leaving them vulnerable to miscalibration, model drift, and subgroup bias. 
These ``silent failures'' can be particularly harmful in high-stakes workflows where errors can delay diagnosis or misdirect care \cite{abikenari2025, fehr2024, nelson2025, muralidharan2024fdaAIreview, smith2022, hall2024, baddam2025}. 
High-profile audits demonstrate how unmonitored algorithms can function as ``silent failures.'' 
An algorithm evaluated by Obermeyer et al. systematically under-referred Black patients for advanced care by using health care costs as a proxy for clinical need, thereby embedding structural bias \cite{obermeyer2019}. 
Similar inequities have been documented in maternal health, where screening tools for perinatal mood and anxiety disorders (PMAD) contribute to undertreatment of minority patients \cite{hall2024, haight2024}. 
Similarly, race-adjusted vaginal birth after cesarean (VBAC) prediction calculators historically underestimated the likelihood of successful trial of labor after cesarean (TOLAC) for Black and Hispanic women, often steering them toward unnecessary cesarean delivery \cite{grobman2007, vyas2020}. 

Early-onset colorectal cancer (EO-CRC), typically defined as colorectal cancer diagnosed before age 50, represents another clinical area marked by wide inequities. In the United States, incidence among younger adults has increased steadily over the past several decades, rising by roughly 3\% per year among individuals aged 20–49 \cite{siegel2017, burnetthartman2021}. 
These trends are accompanied by pronounced disparities in treatment and health outcomes. Racial and ethnic minority populations, particularly Black and Hispanic patients, and individuals from lower socioeconomic backgrounds are more likely to experience delayed diagnosis, present with more advanced disease, and have poorer survival compared with non-Hispanic White patients \cite{murphy2019, kamath2021}. 
Men with early-onset colorectal cancer also demonstrate worse survival outcomes than women \cite{Afify2024}. These disparities reflect a complex interplay of structural and biological factors. 
Differences in access to care, delays in symptom evaluation and treatment, insurance coverage, and neighborhood-level socioeconomic conditions contribute to persistent inequities in stage at diagnosis and receipt of timely care \cite{salem2021, ko2024, loomanskropp2025}. 
In addition, emerging evidence suggests that race- and ethnicity-associated genetic and epigenetic differences may influence tumor biology and disease progression \cite{li2025}.
In this context, where delayed recognition and complex care pathways already shape outcomes, algorithmic tools used for triage, symptom evaluation, referral prioritization, and care navigation may either help reduce inequities through earlier detection or inadvertently amplify them through biased or drifting predictions \cite{obermeyer2019, hall2024}. 
As such, it is imperative to explicitly evaluate and correct for context-specific disparities within clinical AI systems to ensure equitable performance across populations.

At the core of this problem lies a critical structural barrier: an “expertise bottleneck” in AI safety monitoring. Health systems are deploying predictive models at a pace that exceeds their capacity to evaluate safety rigorously and continuously. Algorithmic safety is inherently context-dependent. Determining which performance metrics are most clinically appropriate, such as prioritizing sensitivity in depression or cancer screening versus specificity in TOLAC prediction, requires nuanced clinical judgment and domain expertise. Existing evaluation workflows depend on scarce, cross-disciplinary expert teams to interpret stratified performance metrics and decide on acceptable trade-offs for each new model. This dependence creates three interrelated limitations. First, scalability is constrained; many institutions lack the personnel and infrastructure necessary for comprehensive manual auditing. Second, generic automation tools lack contextual reasoning and cannot distinguish between distinct forms of harm, such as undertreatment in PMAD screening versus overtreatment in VBAC prediction. Third, reliance on manual review introduces substantial latency, with safety assessments often requiring months to complete. During this interval, unmonitored algorithms may continue to generate unsafe or inequitable recommendations.


In this work, we evaluate whether an agentic system can help address gaps in biomedical algorithmic safety auditing by autonomously: 1) identifying relevant sensitive attributes to evaluate a model's performance across (e.g., race, ethnicity, gender, age, etc.) and 2) recommend context-sensitive fairness metrics to evaluate clinical prediction models on.
The system consists of two LLM-driven agents equipped with retrieval-augmented generation (RAG) to extract and synthesize evidence from the clinical literature and identify documented racial, ethnic, and sex- or gender-based disparities. 
We compare the system’s recommendations and supporting rationale against a derived ground-truth solution set.
We study three pretrained LLMs spanning small, medium, and large parameter scales (8B, 20B, and 120B), reflecting the practical constraint that agentic systems may not always have sufficient resources to deploy the largest available models and tools. 
We conduct an ablation study in which each candidate LLM serves as the base model for the agents.
We then evaluate two simplified variants: one without the RAG component, and another without both the agentic orchestration framework and RAG.
This design enables us to isolate and quantify the contributions of RAG and agentic orchestration across models of different scales. 

\section*{Background on Agentic Artificial Intelligence (AI)}

Agentic AI represents an emerging paradigm in intelligent systems \cite{li2026aai}, driven by recent advances in AI, ML, and natural language processing.
Notably, the rapid progress of LLMs built on the Transformer architecture \cite{vaswani2017attention} has revitalized interest in agentic systems, as these models exhibit capabilities that extend beyond conventional chatbot functionality toward more autonomous, human-like task execution.
Broadly, Agentic AI denotes an autonomous computational system that is assigned a human-specified objective and is capable of pursuing that objective with minimal or no ongoing human intervention.
These systems may be instantiated as a single model or as a coordinated ensemble of AI, ML, and LLM tools that collectively enable autonomous reasoning, planning, decision-making, and task execution.
Within biomedicine, agentic methodologies have already been investigated for tackling complex scientific challenges, as demonstrated by systems such as Virtual Lab, Google AI Co-Scientist, and Biomni. Because the primary focus of this work is on Agentic AI systems driven by LLMs, we limit our discussion to these systems.

Although agentic AI systems built on LLMs represent a promising framework for tackling complex research tasks, they continue to inherit several fundamental limitations from the underlying language models. 
One of the most significant challenges is hallucination, in which the system generates outputs that are linguistically plausible but factually or logically incorrect \cite{huang2025survey}.
Moreover, these systems often struggle with long-horizon planning \cite{valmeekam2022large,valmeekam2023planning,tantakoun2025llms}, as their performance may degrade over extended multi-step tasks due to loss of goal alignment, error accumulation, and limited capacity for recovery from earlier failures. 
Effective tool use also remains an open challenge \cite{huang2024metatool}, with agents prone to selecting unsuitable tools, providing invalid inputs, or misunderstanding returned outputs.
In addition, context-window and memory limitations can lead to the loss or dilution of critical instructions, prior observations, and intermediate results as interactions grow in length \cite{liu2024lost}.
These limitations are further compounded by substantial computational cost and latency, because completing a single task may require multiple rounds of reasoning, repeated tool calls, and large context windows, ultimately making deployment less efficient and more resource-intensive.

To evaluate the accuracy of the system's recommended sensitive attributes and fairness metrics, we focus our discussion on reducing hallucinations. Hallucinations in LLMs arise primarily from two factors \cite{tonmoy2024comprehensive}: model architecture and training data.  
First, smaller models may lack sufficient parameters to represent the full range of semantic patterns present in the training corpus.  
Second, when an LLM is not exposed to relevant or domain-specific data during training, it may develop knowledge gaps that impair its ability to answer specialized questions.
Retrieval-Augmented Generation (RAG)\cite{lewis2020retrieval} has emerged as an effective architectural framework for addressing these limitations by integrating the generative capabilities of LLMs with the accuracy and verifiability of external retrieval systems. 
In a typical RAG pipeline, a user query is used to retrieve semantically relevant information from a knowledge source, such as a vector database containing document embeddings. 
These retrieved document chunks are then combined with the original query to form an augmented prompt, which guides the generative model to synthesize a response grounded in specific evidence rather than relying solely on its pre-trained, static parametric memory.
By grounding responses in retrieved source documents, RAG reduces the likelihood of hallucination and enables access to information beyond the model's training process \cite{ayala2024reducing,salemi2024rag}.
As a result, extracted entities, relationships, and other structured data elements can be linked more directly to verifiable and current textual evidence.

\section*{Methods}

The AI safety framework consists of two sequential LLM-driven agents. The first, the \textit{Domain Expert Agent}, generates a structured list of disparity drivers together with a narrative rationale summary. The second, the \textit{Fairness Consultant Agent}, uses this output to recommend fairness metrics and sensitive attributes for model auditing, with a rationale for each recommendation. For both agents, output quality was evaluated by measuring semantic similarity to expert-derived ground truth reference statements using cosine similarity over sentence embeddings. We use semantic similarity as a proxy for recommendation quality rather than a direct measure of factual or clinical correctness.

\begin{figure}[H]
  \centering
  \includegraphics[width=0.85\linewidth]{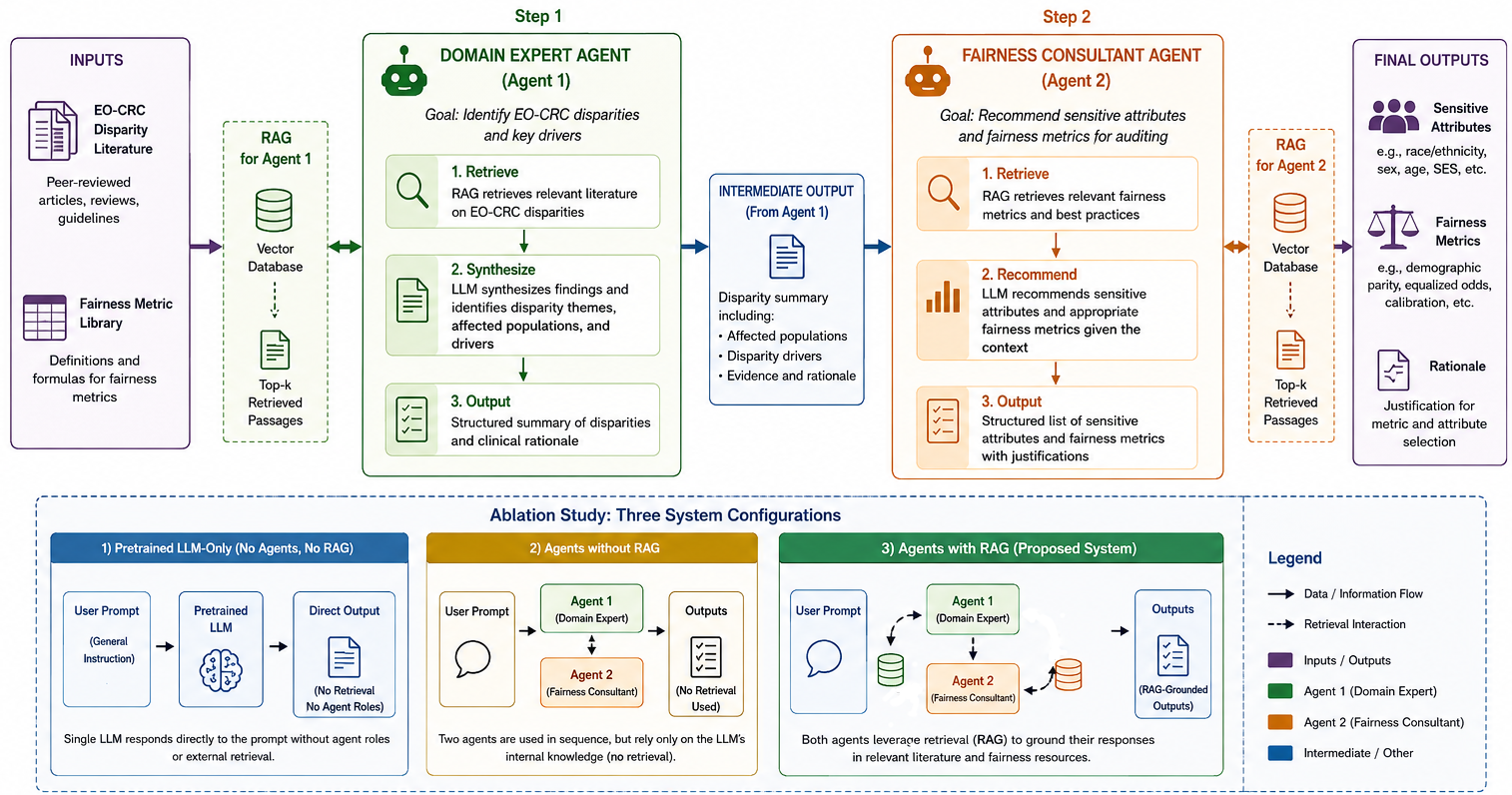}
  \caption{Conceptual model of the agentic fairness auditing framework.}
  \label{fig:flowchart}
\end{figure}

To evaluate the robustness of our agentic-driven system under different configurations, we instantiated the agents with three backbone LLMs: (1) \textbf{Llama 3.1 8B}, (2) \textbf{GPT-OSS 20B}, and (3) \textbf{GPT-OSS 120B}.
This comparison enabled us to assess the extent to which model scale and parameter capacity influence performance, and to determine the minimum model complexity required for reliable completion of the auditing tasks.
We further performed an ablation study comparing the full agentic system with RAG against two reduced variants: (1) a configuration in which each component was replaced by a standalone pretrained LLM, and (2) an LLM-driven agent operating without access to the RAG tool.
Replacing the original RAG-enabled agents with these variants allowed us to isolate the contribution of each component and quantify the improvements provided by the full system. Figure~\ref{fig:flowchart} provides a conceptual overview of the proposed agentic fairness auditing framework.

\textbf{Agent 1: The Domain Expert Agent}. This agent is designed to synthesize clinical literature and identify populations at increased risk. 
The agent utilizes RAG to query a collection of relevant journal articles (e.g., PubMed-indexed journals such as Nature, American Journal of Surgery, and European Journal of Medical Research) related to the clinical context. 
For this study, the corpus comprised 39 articles on EO-CRC: 22 papers explicitly examining disparities in EO-CRC and 17 “noise” papers that discuss EO-CRC more broadly without focusing on health disparities. 

The agent processes these articles and generates a machine-readable JSON object containing two outputs: (1) a structured list of disparity drivers (e.g., socioeconomic status, including lower education levels, poverty, and limited access to health care services; racial and ethnic disparities, with African American and Hispanic populations experiencing higher incidence rates of EO-CRC compared with non-Hispanic White populations), and (2) a narrative summary that synthesizes these factors into a concise clinical rationale. 
Agent 1's performance was evaluated by measuring the semantic similarity between the generated summary and an expert-derived ground truth reference statement. The reference statement was initially drafted by the first author, then reviewed and revised by an EO-CRC researcher, and subsequently by an EO-CRC clinician, who finalized the version used for evaluation (see example in Table~\ref{tab:groundtruth}). 

\textbf{Agent 2: The Fairness Consultant Agent}. This agent utilizes the structured context generated by Agent 1 to execute safety auditing rules. The agent enforces RAG to query an external, literature-based curated Fairness Metric library that contains formal definitions of fairness metrics, such as equalized odds.  
The agent produces a machine-readable JSON summary with three outputs: (1) the fairness metrics recommended for evaluating a predictive model in the given clinical context, (2) the rationale for selecting each metric, and (3) the variables that should be treated as sensitive attributes during model development and evaluation. In EO-CRC, these recommendations emphasize disparities in delayed diagnosis or missed disease. Therefore, metrics that capture inequities in missed positive cases, such as equal opportunity and false negative rate parity, are particularly relevant because underdetection in already underserved groups may worsen downstream outcomes.
Agent 2’s performance was evaluated by measuring the semantic similarity between the generated summary and an expert-derived reference statement, similar to Agent 1's evaluation. The reference statement was initially drafted by the first author and then reviewed by a researcher with expertise in algorithmic fairness (EFW); the final version was used for evaluation (see Table~\ref{tab:groundtruth}). A clinician review step was not conducted for Agent 2's reference statement because the task involves selecting a fairness metric rather than a clinical diagnosis, and a fairness methods expert was considered the appropriate reviewer. These reference statements served as evaluation targets for semantic similarity and were intended to represent expert-reviewed benchmark responses for each task.

\begin{table}[H]
\centering
\caption{Expert-Driven Ground Truth statements used to evaluate performance for Agents 1 and 2.}
\label{tab:groundtruth}
\footnotesize
\setlength{\tabcolsep}{4pt}
\begin{tabularx}{\linewidth}{>{\hsize=0.7\hsize}X >{\hsize=0.3\hsize}X}
\hline
\textbf{Ground Truth Agent 1} & \textbf{Ground Truth Agent 2}\\
\hline

``Yes, there are well-documented disparities in early-onset colorectal cancer in both diagnosis and prognosis. Specifically, racial and ethnic minorities, particularly Black and Hispanic patients, and individuals from lower socioeconomic backgrounds are more likely to experience delayed diagnosis, present with more advanced disease, and have worse survival compared with non-Hispanic White patients. Men with early-onset colorectal cancer also have poorer survival outcomes than women. These disparities are driven by differences in access to care, delays in symptom evaluation and treatment, insurance and neighborhood socioeconomic factors, persistent inequities in stage at diagnosis and receipt of timely care, as well as emerging evidence of race- and ethnicity-associated genetic and epigenetic differences that may influence tumor behavior.'' 
&``Given that the documented disparities lie in missed and delayed diagnoses, and worse survival among Black, Hispanic, lower-SES patients, and men, you should consider including equal opportunity, false negative rate parity, and equalized odds if feasible.'' \\

\hline
\end{tabularx}
\end{table}

\textbf{Ablation Conditions}. 
We evaluated three experimental conditions for each LLM-driven agent configuration with access to the RAG tool: (1) \textbf{pretrained LLM-only}, in which the model performed the task using only its pretraining knowledge; (2) \textbf{Agent without RAG}, in which the agent executed the task using the provided prompt but without retrieving supporting information from external knowledge sources (for Agent 1, the corpus of relevant journal articles; for Agent 2, the fairness metric library); and (3) \textbf{Agent with RAG}, in which the agents performed the task while dynamically retrieving and incorporating relevant information from these external knowledge sources during generation.
Note that each individual condition determines both components of the overarching AI safety framework.

\textbf{LLM \& RAG}. All experiments were executed locally using the Ollama inference framework with three open-source pretrained LLMs: Llama 3.1 (8B), GPT-OSS (20B), and GPT-OSS (120B). 
These models were selected because they are freely available and therefore accessible to a broad range of researchers, aligning with our goal of developing an auditing framework that can be readily adopted by the wider research community.
For experiments using RAG, documents were embedded using the mxbai-embed-large embedding model and indexed in a Chroma vector database for semantic retrieval. 
For Agent 1, retrieval prioritized passages from distinct source articles to increase the diversity of retrieved evidence. All models were run with a temperature of 0.2 to reduce randomness while preserving limited variation across repeated runs.

\section*{Results}

\begin{figure}[H]
  \centering
  \includegraphics[scale=0.3]{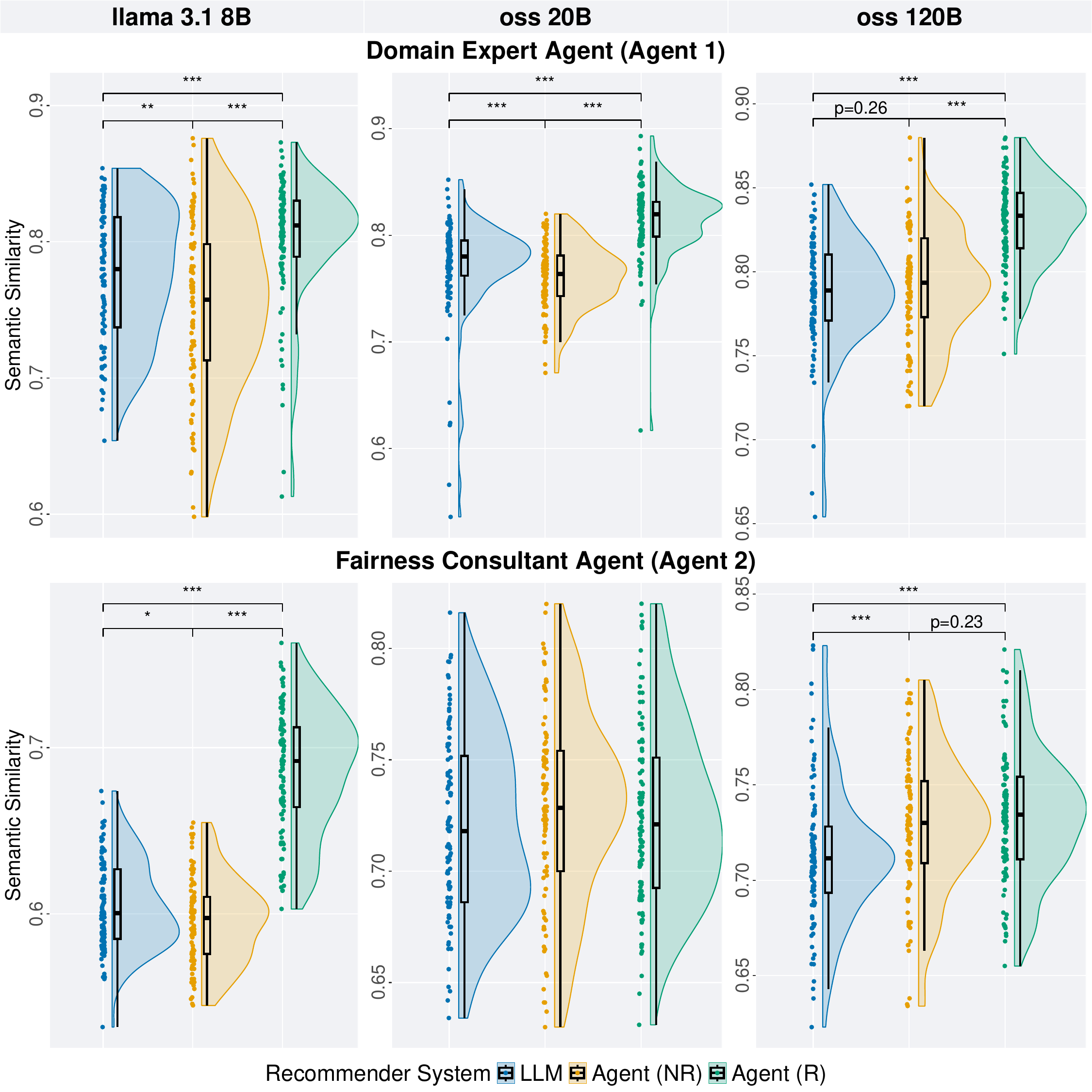}
  \caption{Distribution of semantic similarity scores between output generated by Agents 1 and 2 across three conditions and the relevant ground truth statements. Results are shown for two agent roles: the Domain Expert Agent (Agent 1) and the Fairness Consultant Agent (Agent 2) across three model sizes. Violin plots display the distribution of similarity scores with embedded boxplots and individual observations. Comparisons are shown for the LLM and Agent without RAG (Agent NR) and Agent with RAG (Agent R). Statistical significance of pairwise comparisons is indicated by asterisks, with reported p-values where differences are not statistically significant.}
  \label{fig:plot}
\end{figure}

Semantic similarity between agent outputs and ground-truth statements was evaluated across three system configurations: pretrained LLM-only, Agent without RAG (Agent NR), and Agent with RAG (Agent R) for three models (Llama 3.1 8B, GPT-OSS 20B, and GPT-OSS 120B). Descriptive statistics for each condition are reported in Table~\ref{tab:descriptives}, and inferential test results are summarized in Table~\ref{tab:tests} and Figure~\ref{fig:plot}.

\textbf{Agent 1 Results}.
Across all models, the Agent with RAG configuration consistently produced the highest semantic similarity scores relative to both the pretrained LLM-only and Agent without RAG conditions (see Table~\ref{tab:descriptives}). 
Nonparametric analyses revealed significant differences among conditions for all three models (see Table~\ref{tab:tests}).
Post-hoc Wilcoxon tests with Holm correction indicated that the Agent with RAG condition significantly outperformed the pretrained LLM-only condition across models. 
Additionally, the Agent with RAG condition significantly exceeded the Agent without RAG condition for all models. 
Comparisons between the pretrained LLM-only and Agent without RAG conditions were less consistent, with no significant difference observed for the largest model. 
Overall, these findings indicate that incorporating RAG substantially improves Agent 1’s semantic alignment with the ground-truth statements, and that this benefit persists across models. 
Additionally, we evaluated whether the Agent with RAG condition invoked the retrieval tool during task execution, and found the tool is utilized in 100\% of runs across models, suggesting that the external literature was useful for Agent 1's disparity identification regardless of LLM size.  

\begin{table}[H]
\centering
\caption{Semantic similarity across model scales for Agent 1 and Agent 2 relative to the respective ground truth statements. Values represent mean ($M$), median ($Mdn$), and interquartile range ($IQR$).}
\label{tab:descriptives}

\resizebox{0.6\textwidth}{!}{%
\begin{tabular}{llllccc}
\toprule
Model & Agent & Condition & $n$ & $M$ & $Mdn$ & $IQR$ \\
\midrule

\multirow{6}{*}{\textbf{Llama 3.1 8B}}
& Agent 1 & LLM        & 100 & 0.775 & 0.780 & 0.081 \\
&         & Agent (NR) & 100 & 0.754 & 0.758 & 0.085 \\
&         & Agent (R)  & 100 & 0.802 & 0.812 & 0.041 \\
& Agent 2 & LLM        & 100 & 0.605 & 0.600 & 0.042 \\
&         & Agent (NR) & 100 & 0.596 & 0.597 & 0.034 \\
&         & Agent (R)  & 100 & 0.686 & 0.692 & 0.048 \\
\addlinespace
\hline
\addlinespace
\multirow{6}{*}{\textbf{OSS 20B}}
& Agent 1 & LLM        & 100 & 0.771 & 0.780 & 0.033 \\
&         & Agent (NR) & 100 & 0.761 & 0.764 & 0.038 \\
&         & Agent (R)  & 100 & 0.814 & 0.819 & 0.032 \\
& Agent 2 & LLM        & 100 & 0.720 & 0.718 & 0.066 \\
&         & Agent (NR) & 100 & 0.728 & 0.728 & 0.054 \\
&         & Agent (R)  & 100 & 0.725 & 0.721 & 0.059 \\
\addlinespace
\hline
\addlinespace
\multirow{6}{*}{\textbf{OSS 120B}}
& Agent 1 & LLM        & 100 & 0.788 & 0.789 & 0.039 \\
&         & Agent (NR) & 100 & 0.792 & 0.794 & 0.047 \\
&         & Agent (R)  & 100 & 0.830 & 0.833 & 0.033 \\
& Agent 2 & LLM        & 100 & 0.713 & 0.712 & 0.034 \\
&         & Agent (NR) & 100 & 0.729 & 0.730 & 0.043 \\
&         & Agent (R)  & 100 & 0.733 & 0.734 & 0.043 \\

\bottomrule
\end{tabular}
}
\end{table}

\textbf{Agent 2 Results}.
Results for Agent 2 showed a more variable pattern across models (see Table~\ref{tab:descriptives} and \ref{tab:tests}).
For the Llama 3.1 8B model, significant differences among conditions were observed, with the Agent with RAG configuration producing higher semantic similarity than the other conditions. 
For the OSS 20B model, however, the omnibus test did not detect significant differences among conditions, indicating that system configuration had little effect on Agent 2 performance at this scale. 
In contrast, significant differences re-emerged for the OSS 120B model, where both agent-based configurations yielded higher semantic similarity scores than the pretrained LLM-only condition. 
Pairwise comparisons indicated that the Agent with RAG configuration significantly exceeded the pretrained LLM-only baseline, although differences between the Agent with RAG and without RAG were not significant. 
When evaluating whether the Agent with RAG invoked the retrieval tool, we found 100\% tool use across models, except for the GPT-OSS 120B model, which showed an 85\% retrieval tool use for Agent 2. 
These findings indicate that the agent architecture reliably triggered retrieval behavior across models, though the largest model occasionally completed the task without invoking external knowledge.



\begin{table}[H]
\centering
\caption{Statistical tests for semantic similarity across model scales. Overall differences were evaluated using Kruskal--Wallis tests. Pairwise comparisons were conducted using Wilcoxon rank-sum tests with Holm correction.}
\label{tab:tests}

\resizebox{0.7\textwidth}{!}{%
\begin{tabular}{llll}
\toprule
Model & Agent & Overall test & Pairwise comparison ($p$ Holm) \\
\midrule

\multirow{6}{*}{\textbf{Llama 3.1 8B}}
& Agent 1 & $H(2)=39.56$, $p<.001$ & Agent (NR) $<$ LLM ($p=.007$) \\
&         &                        & Agent (R) $>$ LLM ($p<.001$) \\
&         &                        & Agent (R) $>$ Agent (NR) ($p<.001$) \\
\addlinespace
& Agent 2 & $H(2)=173.73$, $p<.001$ & Agent (NR) $<$ LLM ($p=.014$) \\
&         &                         & Agent (R) $>$ LLM ($p<.001$) \\
&         &                         & Agent (R) $>$ Agent (NR) ($p<.001$) \\

\addlinespace
\hline
\addlinespace\multirow{6}{*}{\textbf{OSS 20B}}
& Agent 1 & $H(2)=114.45$, $p<.001$ & Agent (NR) $<$ LLM ($p<.001$) \\
&         &                         & Agent (R) $>$ LLM ($p<.001$) \\
&         &                         & Agent (R) $>$ Agent (NR) ($p<.001$) \\
\addlinespace
& Agent 2 & $H(2)=2.38$, $p=.305$ & No significant differences observed \\
&         &                       & Pairwise tests not interpreted \\
&         &                       & -- \\

\addlinespace
\hline
\addlinespace\multirow{6}{*}{\textbf{OSS 120B}}
& Agent 1 & $H(2)=93.53$, $p<.001$ & Agent (NR) $>$ LLM ($p=.26$) \\
&         &                        & Agent (R) $>$ LLM ($p<.001$) \\
&         &                        & Agent (R) $>$ Agent (NR) ($p<.001$) \\
\addlinespace
& Agent 2 & $H(2)=20.77$, $p<.001$ & Agent (NR) $>$ LLM ($p<.001$) \\
&         &                        & Agent (R) $>$ LLM ($p<.001$) \\
&         &                        & Agent (R) $>$ Agent (NR) ($p=.24$) \\

\bottomrule
\end{tabular}
}
\end{table}

\section*{Discussion}

The present study examined how RAG and agent architecture influence semantic alignment between system outputs and expert-generated ground-truth statements across three model scales. 
For Agent 1, retrieval consistently improved semantic alignment across all models, suggesting that grounding responses in external evidence enhances performance for disparity-identification tasks (see Table~\ref{tab:descriptives} and \ref{tab:tests}).  In contrast, Agent 2 demonstrated more variable improvements across models, suggesting that the benefits of retrieval may depend more on model scale for tasks involving fairness metric selection and sensitive attribute recommendation.

One possible explanation is that the two agents serve distinct functional roles within the overall architecture and therefore rely on different types of knowledge. Agent 1 depends on specialized knowledge about disparities in early-onset colorectal cancer and thus benefits substantially from grounding in external evidence. In contrast, Agent 2 relies more heavily on internal reasoning and synthesis of fairness concepts that may already be represented in the model’s pretraining, reducing the relative importance of retrieval. This distinction may explain why retrieval produced stronger and more consistent performance gains for Agent 1 than for Agent 2, even though Agent 2 invoked retrieval in most runs.


These varied findings may reflect that LLMs already possess prior knowledge of fairness metrics from their baseline training. As a result, the models may rely less on externally retrieved information when performing tasks involving fairness reasoning. Consistent with this interpretation, even when Agent 2 was explicitly prompted to use retrieval-augmented generation, the largest model evaluated in this study invoked the retrieval tool in approximately 85\% of runs. This pattern suggests that the model’s internal knowledge may have been sufficient to complete the task in a substantial proportion of cases.

The current implementation relies on literature-based retrieval rather than patient-level EHR data or live clinical workflows, which represents an important consideration for real-world translation. In practice, deployment within a health system would require integration with institutional data infrastructure, including structured EHR outputs, model performance logs, and potentially real-time prediction feeds. Such integration would enable the system to move beyond recommending sensitive attributes and fairness metrics in the abstract toward auditing specific deployed models against observed subgroup performance. While the present study establishes proof-of-concept for the agentic auditing approach, future work should evaluate the system in a clinical deployment context. For example, by connecting Agent 2's fairness metric recommendations directly to stratified performance reports generated from a live or retrospective EHR cohort. Partnerships with health system informatics teams and attention to data governance, privacy, and institutional review will be essential for such extensions.

Several limitations should be considered. First, semantic similarity to expert-derived ground-truth statements serves only as a proxy for recommendation quality; it does not directly assess factual correctness, reasoning validity, or clinical appropriateness. Dedicated hallucination benchmarks, such as those assessing factual grounding against source documents, represent a more direct evaluation strategy and should be incorporated in future work. Second, the analysis examined a limited set of LLMs and a single clinical domain. Future work should include direct expert assessment of output quality, broader model comparisons, agent interactions, and additional clinical domains.

\section*{Conclusion}

Overall, the findings suggest that RAG can improve semantic alignment between agent-generated outputs and expert-derived reference standards, especially for tasks that require grounding in external knowledge sources. These benefits were strongest and most consistent for Agent 1, whereas gains for Agent 2 were more variable across model scales, highlighting the importance of considering both task type and model capacity when developing agent-based AI systems. 



In resource-constrained settings, the Llama 3.1 8B model may be a practical option when paired with an agentic framework that can effectively leverage a RAG tool with access to up-to-date knowledge. From a computational perspective, this model is also more feasible to run on local hardware (e.g., a sufficiently capable laptop), than the larger models evaluated in this study. Future work should investigate a broader landscape of LLMs operating under similar hardware constraints to systematically assess performance in these environments.

In settings with greater computational resources, the GPT-OSS 120B model showed the strongest semantic-similarity performance in this study when used with an agentic framework that integrates a RAG tool providing access to current knowledge. 
This is evident in the model's largest mean and median across the configurations assessed in this study (Table \ref{tab:descriptives}). Future work should compare a broader range of LLMs within the same agentic framework and retrieval setup to determine whether sufficiently large models yield meaningful performance differences.

More broadly, continued investigation in this area may help identify which combinations of model scale, retrieval, and agent structure  are most effective for generating expert-aligned auditing recommendations. It may also help reveal smaller models that remain capable while being accessible on modest hardware. Such progress could improve the accessibility of agentic auditing systems and support more equitable AI development in healthcare.

\renewcommand{\refname}{\hfill References\hfill}
\bibliography{references}

\end{document}